\documentclass[reprint,superscriptaddress,showpacs,showkeys,amsmath,amssymb,floatfix,aps,pra,longbibliography]{revtex4-1}

\usepackage{graphicx}
\usepackage{bm}

\usepackage[dvipsnames]{xcolor}
\usepackage{physics}
\usepackage{placeins}
\usepackage{graphics}
\usepackage{color}
\usepackage{hyperref}
\usepackage{multirow}
\usepackage{blindtext}
\usepackage{gensymb}
\usepackage{textcomp}
\begin{document}


\title{Twist-angle dependent proximity induced spin-orbit coupling\\ in graphene/transition-metal dichalcogenide heterostructures}

\author{Thomas Naimer}
\email{thomas.naimer@physik.uni-regensburg.de}
\affiliation{Institute for Theoretical Physics, University of Regensburg, 93040 Regensburg, Germany}
\author{Klaus Zollner}
\affiliation{Institute for Theoretical Physics, University of Regensburg, 93040 Regensburg, Germany}
\author{Martin Gmitra}
\affiliation{Institute of Physics, Pavol Jozef Šafárik University in Košice, 04001 Košice, Slovakia}
\author{Jaroslav Fabian}
\affiliation{Institute for Theoretical Physics, University of Regensburg, 93040 Regensburg, Germany}

\begin{abstract}
We investigate the proximity-induced spin-orbit coupling in heterostructures of twisted graphene and monolayers of transition-metal dichalcogenides (TMDCs) MoS$_2$, WS$_2$, MoSe$_2$, and WSe$_2$ from first principles. We identify strain, which is necessary to define commensurate supercells, as the key factor affecting the band offsets and thus magnitudes of the proximity couplings. We establish that for biaxially strained graphene the band offsets between the Dirac point and conduction (valence) TMDC bands vary linearly with strain, regardless of the twist angle. This relation allows to identify the apparent zero-strain band offsets and find a compensating transverse electric field correcting for the strain. The resulting corrected band structure is then fitted around the Dirac point to an established spin-orbit Hamiltonian. This procedure yields the dominant, valley-Zeeman and Rashba spin-orbit couplings. The magnitudes of these couplings do not vary much with  the twist angle, although the valley-Zeeman coupling vanishes for 30$\degree$~and Mo-based heterostructures exhibit a maximum of the coupling at around 20$\degree$. The maximum for W-based stacks is at 0$\degree$.  The Rashba coupling is in general weaker than the valley-Zeeman coupling, except at angles close to 30$\degree$. We also identify the Rashba phase angle which measures the deviation of the in-plane spin texture from tangential, and find that this angle is very sensitive to the applied transverse electric field. 
We further discuss the reliability of the supercell approach with respect to atomic relaxation (rippling of graphene), relative lateral shifts of the atomic layers, and transverse electric field.   
\end{abstract}

\pacs{}
\keywords{spintronics, graphene, TMDC, heterostructures, proximity spin-orbit coupling}
\maketitle

\section{Introduction}

Graphene's weak (tens of $\mu$eV) spin-orbit coupling (SOC)~\cite{Gmitra2009:GrapheneEfield,Sichau2019:PRL} enables long spin diffusion paths but precludes it from 
forming a platform for spin manipulation~\citep{Han2014:NN}. Fortunately, proximity effects enable strong SOC---on the meV scale---in graphene~\citep{Sierra2021}, making it a suitable candidate for spintronics devices~\cite{Zutic2004:RMP,Avsar2019:arxiv}
and for observing topological states
~\citep{Kane2005:PRL,Kane2005:PRL2,Qiao2010:PRB,Ren2016:RPP,Frank2018:pseudohelical,Hogl2020:PRL}.

Unlike pristine graphene, whose spin-orbit coupling is of the Kane-Mele type~\cite{Kane2005:PRL,Sichau2019:PRL,Gmitra2009:GrapheneEfield}, graphene proximitized by
transition-metal dichalcogenides (TMDCs) 
exhibits valley-Zeeman SOC which acts as an
effective magnetic field, opposite at $K$ and $K'$, but otherwise momentum independent close to these points~\citep{Gmitra2015:TMDCgraphene1, Wang2015:NC,Gmitra2016:TMDCgraphene2,Frank2018:pseudohelical}; similarly for graphene on topological insulators~\cite{Song2018:TIGRheteroDFT,Zollner2019b:PRB,Zollner2021:PSSB,Khokhriakov2018:TIGRheteroexp}. There is by now plenty of experimental evidence for the large proximity SOC in graphene~\citep{Avsar2014:NC,Wang2015:NC,Yang2016:2DM,Wang2016:PRX,Voelkl2017:proxSOCexp,Zihlmann2018:PRB,Khokhriakov2018:TIGRheteroexp,Garcia2018:CSR,Safeer2019:NL,Herlin2020:APL,Khoo2017:NL,Wang2015:NC,Omar2018:PRB,Omar2017:PRB,Fulop2021}. Ramifications of the proximity effect include a spin-orbit valve~\citep{Gmitra2017:SOValve1,Island2019:SOValve2,Tiwari2021:SOValve3,Amann2021:arxiv,Khoo2017:NL} and giant spin-relaxation anisotropies~\citep{Cummings2017:anisotropy1,Ghiasi2017:anisotropy2,Benitez2018:anitrosopy3,Zihlmann2018:PRB}.

The discovery of superconductivity in magic-angle twisted bilayer graphene~\citep{Cao2018:magicangleBLGorig,Arora2020:arxiv,Stepanov2020:Nat,Balents2020:NP} gave the impetus for twistronics~\cite{Carr2017:PRB,Hennighausen2021:ES,Ribeiro2018:SC,Carr2020:NRM}, recognizing the twist angle as
a critical new control parameter for tailoring electronic properties of van der Waals heterostructures~\cite{Zollner2021:arXiv,Yan2021:PE,David2019:TwistTB2,Li2019:TwistTB1,Pezo2021:TwistDFT4}. It is then natural to ask how does the twist angle influence proximity SOC in graphene. Two recent pioneering studies~\citep{Li2019:TwistTB1, David2019:TwistTB2}  based on tight-binding modeling~\citep{Koshino2015:TwistTBBasic} 
suggest that the twist angle between the TMDC and graphene has a major influence on both the magnitude and the type of proximity induced SOC. In those works, the interlayer interaction in incommensurate systems is described by Umklapp processes which connect selected points of the TMDC Brillouin zone with the $K$ and $K'$ points of the graphene Brillouin zone. As these momentum points vary with the twist angle, the proximity SOC gets modified. The reliability of such calculations depends on how well both the energy dispersion of the TMDC layer and the interlayer hybridization between graphene and TMDC orbitals are described by the effective microscopic models. 

Perhaps the most direct theoretical approach for extracting 
proximity SOC in graphene/TMDC bilayers is to perform \textit{ab initio} calculations on supercell geometries incorporating the twisted monolayers. Since density functional theory (DFT) calculations in this regard have been performed almost exclusively on aligned heterostructures (0\degree~twist angle), there is yet no clear \textit{ab initio} perspective on this topic. There already exist studies ~\citep{Wang2015:NC,Hou2017:TwistDFT2,DiFelice2017:TwistDFT3} discussing electronic properties of twisted graphene/TMDC stacks, but not examining the proximity SOC. Recently though, Pezo \textit{et al.}~\citep{Pezo2021:TwistDFT4} reported SOC parameters from DFT calculations on three twist angles (0\degree, 15\degree,~and 30\degree) for graphene/MoTe$_2$ and graphene/WSe$_2$, and examined ramifications for spin relaxation anisotropy.

Here, we aim to provide a comprehensive picture, applying DFT to large graphene/TMDC heterostructures employing 
four different semiconducting TMDCs MoS$_2$, MoSe$_2$, WSe$_2$, and WS$_2$. For several
twist angles we build commensurate supercells with built-in strains in graphene (TMDCs are left unstrained) of up to 17\%, 
although we deem as excessive strains of more than 10\%. 
From the band structures of the supercells we find that the apparent dependence of the band offsets between the Dirac point and the TMDC bands is due to the strain. In other words, heterostructures of different twist angles but similar strain exhibit similar band offsets. By plotting the band offsets as a function of the strain we find a scattered linear relation which allows us to extract the ``zero-strain" offset. To correct for the strain we then apply a transverse electric field, different for different heterostructures, so that at the end we compare different extracted parameters for the same band offsets. We believe that this currently is the best way to extract quantitatively reliable twist-angle dependence of the proximity spin-orbit parameters for computationally feasible supercell sizes. We compare our results with previous tight-binding studies~\citep{Li2019:TwistTB1,David2019:TwistTB2}.

Overall, the magnitude of the SOC parameters are on the meV scale, while the Rashba coupling is typically weaker than the dominant valley-Zeeman coupling. Only close to 30$\degree$, at which the valley-Zeeman coupling vanishes due to symmetry, does Rashba coupling prevail. There is also a marked difference for Mo- and W-based heterostructures. For Mo-based stacks the valley-Zeeman coupling exhibits a global maximum at a peak around 20$\degree$, reaching a value twice that at 0$\degree$. In contrast, while there is a local maximum for W-based stacks close in the region of 15$\degree-20\degree$, the largest value of the valley-Zeeman coupling is found at 0$\degree$. The largest parameters for W-based structures are roughly twice as large as the largest ones for Mo-structures. 

Unlike at 0$\degree$ and 30$\degree$, at a general twist angle between the two extremes (outside of this interval the parameters can be obtained from symmetry arguments as presented below) the Rashba coupling exhibits a radial component which results in a spin texture which deviates from the typical tangential Rashba pattern. The deviation is measured by the Rashba phase angle~\cite{Li2019:TwistTB1,David2019:TwistTB2}.
We also report these angles for selected supercells and demonstrate their rather high sensitivity to the electric field, but not to lateral shifts of the layers. 

There certainly are still many pitfalls of the methodology we use. Finite supercells are susceptible to the atomic registry and atomic relaxation leading to graphene rippling. We also investigate the dependence of the proximity spin-orbit coupling parameters on such effects, although in a limited way due to computational complexity, focusing on selected heterostructures to provide realistic expectations rather than sweeping proofs. For example, we find that lateral shifts of the atomic layers do not change significantly the spin-orbit parameters, while the rippling of graphene strongly enhances the staggered potential and increases the Kane-Mele coupling, though valley-Zeeman coupling still dominates. Also, the compensating electric field to correct the band offsets is not the panacea.
Especially for excessive strain above 10\% the electric field can modify the band structure itself and induce Rashba spin-orbit couplings that are not native to the heterostructures. We provide a discussion of the electric field effects as well.

The paper is organized as follows. In Sec.~\ref{Sec:geo} we introduce supercell geometries, the computational methods are detailed in App.~\ref{App:A}.
The influence of strain on the band offsets and our method of correcting them is presented in Sec.~\ref{Sec:offsetcorr}.
In Sec.~\ref{Sec:effHam} we introduce the effective low energy Dirac Hamiltonian, which is used to extract orbital and proximity SOC parameters.
Finally, in Sec.~\ref{Sec:Results} we discuss the extracted parameters for different angles. We additionally comment on the effect of lateral shifts and structural relaxation in App.~\ref{App:shift} and App.~\ref{App:relax} respectively. In App.~\ref{App:imagRashba} we quantify the Rashba phase angle for different TMDCs and twist angles. 
Using a specific example, we demonstrate in App.~\ref{App:Efield} how a transverse electric field tunes the extracted spin-orbit parameters.

\section{Supercell geometries}
\label{Sec:geo}

Starting from the primitive hexagonal unit cells of graphene and four TMDCs (MoS$_2$, MoSe$_2$, WSe$_2$ and WS$_2$; geometry parameters listed in Tab.~\ref{Tab:latconsts}), we construct the supercells by implementing the coincidence lattice method~\citep{Koda2016:JPCC,Carr2020:NRM, Wang2015:Gut}. Using the integers $n$ and $m$, we define the lattice vectors $\mathbf{a}^S_{(n,m)}$ and $\mathbf{b}^S_{(n,m)}$ of a new supercell as a linear combination of the primitive lattice vectors $\mathbf{a}$ and $\mathbf{b}$:
\begin{align}
\mathbf{a}^S_{(n,m)}&=n\cdot \mathbf{a}+m\cdot \mathbf{b} \\
\mathbf{b}^S_{(n,m)}&=-m\cdot \mathbf{a}+(n+m)\cdot \mathbf{b}.
\end{align}
We give attributes $(n,m)$ to such a supercell. Its lattice constant is
\begin{align}
a^{S}_{(n,m)}=a\cdot \sqrt{(n^2+m^2)+n\cdot m}  ,  
\end{align} 
where $a= |\mathbf{a}|=|\mathbf{b}|$ is the lattice constant of the primitive unit cell.
The relative twist angle with respect to the primitive unit cell is given by
\begin{align}
\Theta_{(n,m)}=\arctan(\frac{\sqrt{3}m}{2n+m}) .
\end{align}
A graphene/TMDC heterostructure supercell contains a ($n$,$m$) graphene supercell beneath a ($n'$,$m'$) TMDC supercell resulting in a relative twist angle 
\begin{align*}
\Theta=& \Theta_{(n,m)}-\Theta_{(n',m')}\\
=& \arctan\Big(\frac{\sqrt{3}m}{2n+m}\Big)-\arctan\Big(\frac{\sqrt{3}m'}{2n'+m'}\Big). 
\end{align*}

    \begin{table}[htb]
    \caption{\label{Tab:Structures} Structural information of the investigated graphene/TMDC heterostructures. 
    We list the supercell attributes $(n,m)$ of graphene and $(n',m')$ of the TMDC. We also list the absolute value of the twist angle $|\Theta|$ between the two monolayers (the sign of $\Theta$ is not relevant here), the strain $\epsilon^{\text{TMDC}}$ in graphene (which depends on the specific choice of TMDC) and the number of atoms ($N_{\text{at}}$) in the heterostructure. For completeness, we also list strains corresponding to supercells, which were not investigated, in grey - for a full list of the supercells used in Fig.~\ref{Fig:params} and Fig.~\ref{Fig:strainVSoffset}, see Tab.~\ref{Tab:param}. These supercells, which were not investigated, either had too much built-in strain, too many atoms or entailed computational difficulties (e.g., convergence problems).}
   
    \begin{ruledtabular}
    \begin{tabular}{cccccccc}

$|\Theta |$ &$(n,m)$ & $(n',m')$& $\epsilon^{\text{MoS}_2}$ & $\epsilon^{\text{WS}_2}$& $\epsilon^{\text{MoSe}_2}$& $\epsilon^{\text{WSe}_2}$&$N_\text{at}$ \\
$[\degree]$&&&[\%]&[\%]&[\%]&[\%]&\\
\hline

0.0      &(4,0)        &(3,0)      & {-2.9} & -3.05 & 1.19 & 1.19        & 59 \\
0.0      &(0,5)        &(0,4)      & 3.58 & 3.41 & 7.93 & 7.93        & 98\\
5.2      &(3,1)        &(2,1)      & -4.99 & -5.14 & -1.0  & -1.0        & 47  \\
6.6      &(3,2)        &(2,2)     & 2.89 & \color{gray}2.73 & 7.22& 7.22        & 74  \\
9.5      &(3,2)        &(3,1)     & 7.1 & \color{gray}6.93 & 11.6& \color{gray}11.6        & 77  \\
10.9     &(2,1)        &(1,1)     &  \color{gray}-15.24 & \color{gray}-15.37 & -11.67& -11.67        & 23  \\
13.9     &(3,1)        &(3,0)     & 7.73 & 7.56 & 12.26& \color{gray}12.26        & 53  \\
13.9     &(0,4)        &(1,3)       & 16.7 & \color{gray}16.52 & \color{gray}21.61& \color{gray}21.61        & 71  \\
13.9     &(5,0)        &(3,1)     & -6.64 & \color{gray}-6.78 & -2.71& -2.71        & 89  \\
19.1     &(2,1)        &(2,0)     & -2.13 & -2.28 & 1.99& 1.99        & 26  \\
22.7     &(3,2)        &(1,3)      & 7.1 & \color{gray}6.93 & 11.6&\color{gray} 11.6        & 77  \\
23.4     &(3,2)        &(3,0)      & \color{gray}-10.89 & \color{gray}-11.03 & -7.14& -7.14        & 65  \\
27.0     &(3,1)        &(1,2)       & -4.99 & -5.14 & -1.0 & -1.0        & 47  \\
30.0     &(2,0)        &(1,1)     & 12.13 & 11.95 & 16.84& 16.84        & 17 \\
30.0     &(5,0)        &(2,2)     & -10.3 & \color{gray}-10.44 & -6.53 & \color{gray}-6.53        & 86  \\

    \end{tabular}
    \end{ruledtabular}
    \end{table}
    \begin{table}[htb]
    \caption{Unstrained geometries of the primitive unit cells of graphene and four selected TMDCs~\citep{Zollner2019:TMDClatconst}. The structure of the TMDCs stay unchanged in the supercells, while the graphene layers are strained by the factors listed in Tab.~\ref{Tab:Structures} to ensure commensurability.}\label{Tab:latconsts}
        \begin{ruledtabular}
        \begin{tabular}{c|ccccc}

				& Graphene 	&  MoS$_2$ &WS$_2$	&MoSe$_2$ 	&WSe$_2$ 	\\
				\hline
$a$[\AA] 		&	2.46 	& 3.185 	&3.18   &	3.319	&	3.319 \\
$d_{\text{XX}}$[\AA] 	&	-		& 3.138		& 3.145 &   3.357	&   3.364		
     	\end{tabular}
        \end{ruledtabular}

    \end{table}

    \begin{figure}[htb]
     \includegraphics[width=.99\columnwidth]{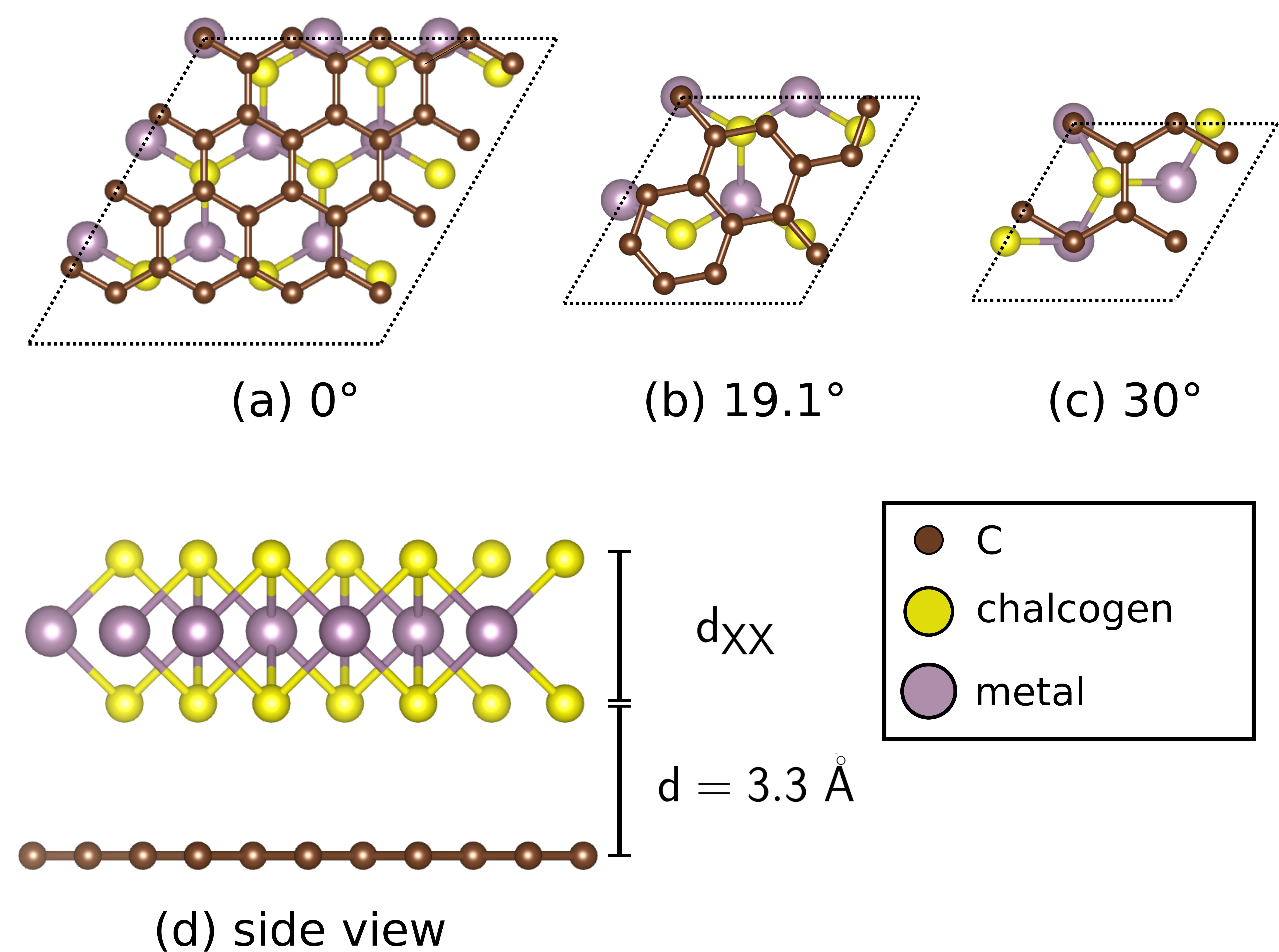}
     \caption{(a)-(c) Bottom view of the graphene/TMDC heterostructure supercells with twist angles $|\Theta|= 0\degree,19.1\degree,30\degree$ (d) Side view of the 0\degree~unit cell with indicated interlayer distance $d$ and chalcogen-chalcogen distance $d_{\text{XX}}$.}\label{Fig:Structures}
    \end{figure}

In our supercells there is at a corner a carbon atom directly beneath a metal atom of the TMDC, see Fig.~\ref{Fig:Structures}. Considering different configurations (see App.~\ref{App:shift}) we find that proximity SOC is rather insensitive to the changes of the atomic registry, as was already shown for $0\degree$  in Ref.~\cite{Gmitra2016:TMDCgraphene2}.
In order to obtain commensurate supercells for periodic DFT calculations, one of the layers (or both) need to be strained. We thus
introduce the strain factor  $\epsilon$ which depends on the twist angle and the lattice constant of the TMDC. TMDCs are very sensitive to strain~\citep{Zollner2019:TMDClatconst}, while the low energy Dirac spectrum of graphene is---apart from a renormalization of the Fermi 
velocity---rather robust against biaxial strain smaller than 20\%~\citep{Si2016:graphenestrain1,Choi2010:graphenestrain2}. Therefore, we choose to leave the TMDC unstrained and instead strain graphene. 
Also, 
we use the same interlayer distance $d=3.3$~\AA~ separating the monolayers (see App.~\ref{App:relax}) for all studied 
supercells, to focus on twist-angle effects. However, in App.~\ref{App:relax} we discuss the effects of structural relaxation and rippling on the proximity band structure at the Dirac cone. 
Finally, we add a vacuum of 20~\AA~to avoid interactions between periodic images in our slab geometry. The graphene/TMDC heterostructures are set up using the {\tt atomic simulation environment (ASE)}~\citep{ASE} code.
The structural parameters of the
heterostructures are collected in  Tab.~\ref{Tab:Structures} and some prominent examples are visualized in Fig.~\ref{Fig:Structures}. 
    
When a hexagonal system is described by a $(n,m)$ supercell, the K point of the Brillouin zone can fold back to either  K, K', or $\Gamma$ of the reduced Brillouin zone. The following rule, which can be derived from geometrical considerations, determines, which of the three options is the case:
\begin{align}
\text{backfolding to $\Gamma$ for:  } n-m=0+3\cdot l \\
\text{backfolding to K for: } n-m=1+3\cdot l\\
\text{backfolding to K' for: } n-m=2+3\cdot l
\end{align}
with $l\in\mathbb{Z}$.
Because of these backfolding effects, the TMDC band gap resides at the K- or the $\Gamma$-point depending on the exact supercell. The Dirac cone can in principle also fold back to $\Gamma$, e.g., for a $(n,m)=(1,1)$ supercell (i.e., a $\sqrt{3}\cross\sqrt{3}$ supercell).
In the following we only consider supercells for which the Dirac cone folds back to the K/K'-point to be able to compare different structures and  extract the spin-orbit parameters using our model Hamiltonian presented below.

Computational methodology for obtaining DFT electronic band structures of the supercells is detailed in App.~\ref{App:A}.

\section{Correcting band offsets for strain}
\label{Sec:offsetcorr}

For most investigated twisted heterostructures we find that the Dirac cone of graphene lies within the TMDC band gap for most supercells. Exceptions are stacks with heavily strained graphene.

    \begin{figure}[htb]
     \includegraphics[width=.75\linewidth]{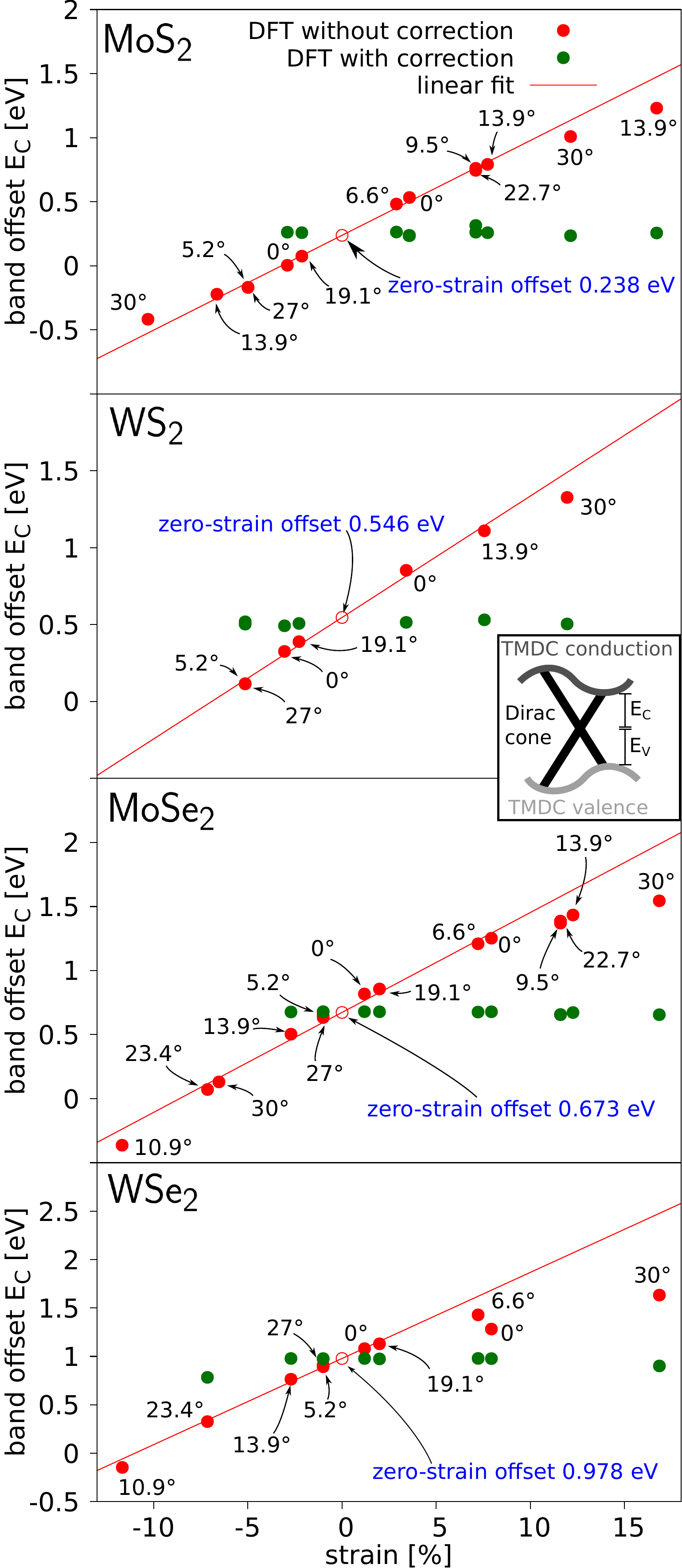} 
     \caption{Correcting for the strain. For all the investigated supercells of graphene with MoS$_2$, WS$_2$, MoSe$_2$, and WSe$_2$ monolayers we plot the band offsets $E_C$ of the Dirac cone with respect to conduction band  (see inset) against the strain on graphene $\epsilon$; $\epsilon >0$ indicates tensile strain while $\epsilon<0$  indicates compressive strain. Each of the data points (red solid circles) is annotated with the twist angle of the corresponding supercell.  From the linear fit (red line) we extract the (apparent) zero-strain band offsets (empty red circles) which are collected in Tab.~\ref{Tab:zerostrainoffsets}. The green circles show the band offsets after the correction by the transverse electric field employed to compensate the influence of strain. Strains above 10$\%$ and below -10$\%$ and negative band offsets $E_C < 0$ were not included in the linear fit.}\label{Fig:strainVSoffset}
    \end{figure}

    \begin{figure*}[htb]
     \includegraphics[width=.85\linewidth]{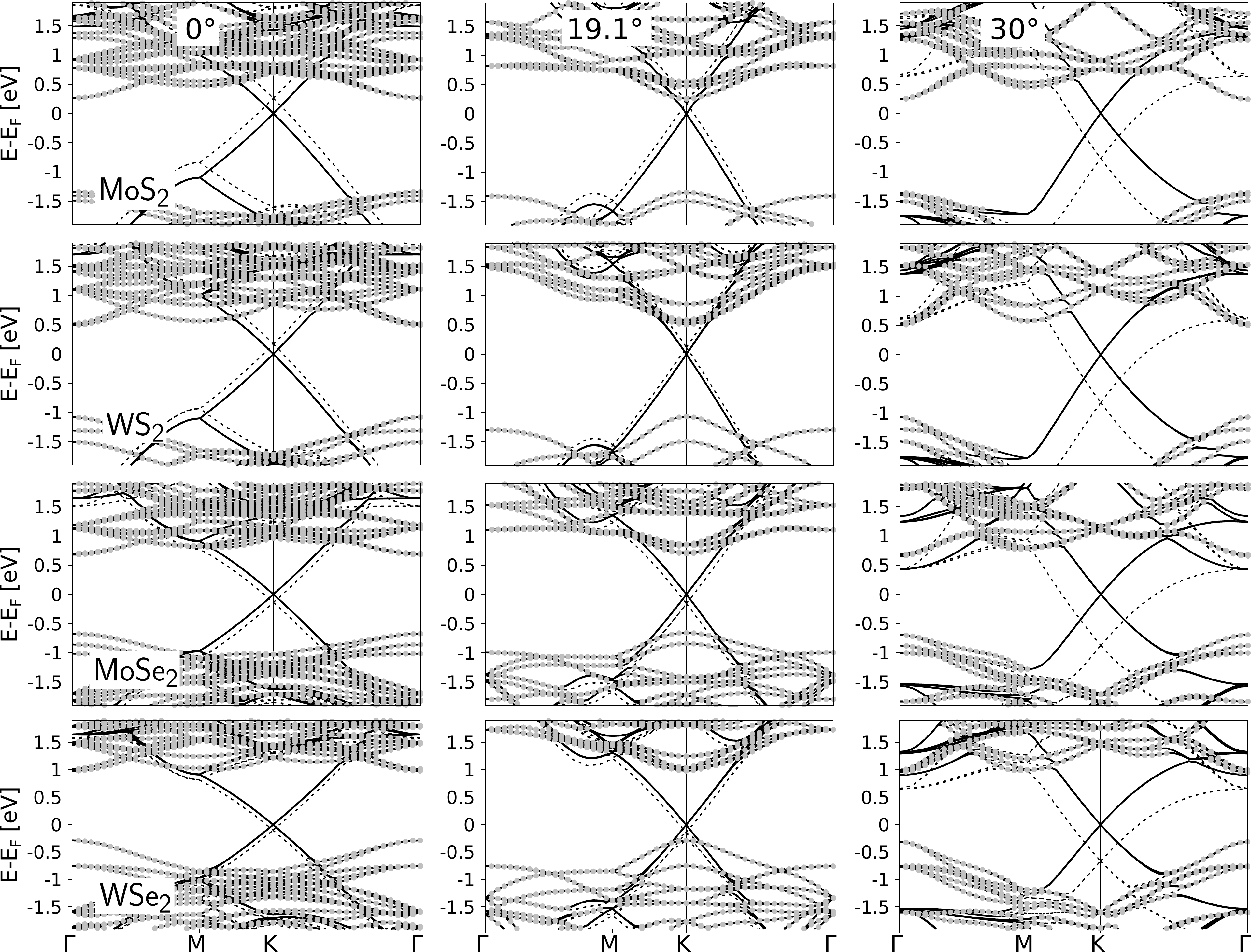} 
     \caption{Calculated band structures along high symmetry lines for selected graphene/TMDC supercells. For each of the four TMDCs, we show three different angles: $|\Theta|= 0\degree,19.1\degree,30\degree$, as indicated. The grey circles originate from TMDC orbitals, while the solid (dashed) lines come from graphene states with (without) the electric field corrections. The energy scale is set to zero for the Fermi energy of the electric-field corrected band structures. Because of band folding effects, the TMDC bands look different, while the global TMDC band gap stays intact for different twist angles.}\label{Fig:GlobalBS}
    \end{figure*}
	
Crucially, the relative twist angle $\Theta$ barely influences the band offset between the Dirac point and the TMDC conduction (valence) band minimum (maximum). Instead, we trace the
considerable band offset variations among the investigated heterostructures to the applied strain which is otherwise necessary to build commensurate supercells of manageable sizes.
In Fig.~\ref{Fig:strainVSoffset} we collect the data for the investigated structures and plot the band offset $E_C$ to the conduction band as a function of the strain $\epsilon$ applied to graphene. The band offset varies, with some scatter,  linearly with $\epsilon$. The above thesis, that it is the strain and not twist that causes the band offset variation, is most strikingly seen in the data of disparate twist angles and very similar strain, such as the three angles 9.5$\degree$, 22.7$\degree$, and 13.9$\degree$ for graphene/MoS$_2$, see Fig.~\ref{Fig:strainVSoffset}. These three heterostructures feature essentially the same offset $E_C$.

The absolute deformation potential of the Dirac cone can be defined as the change of its energy when applying strain. Using the TMDCs band edges (which remain at the same energy, because we leave the TMDC unstrained) as a fix point, we can identify $|\alpha_{D}|$ as the absolute value of the slope of the linear fits in Fig.~\ref{Fig:strainVSoffset}:
\begin{align}
\alpha_{D}=\frac{\partial E_{D}}{\partial\epsilon}=\frac{\partial E_{V}}{\partial\epsilon}=-\frac{\partial E_{C}}{\partial\epsilon} \label{Eq:absdefpot}
\end{align}
Here $E_D$ is the energy level of the Dirac cone and $E_V$ and $E_C$ are the band offsets to the valence and conduction band respectively. We find $\alpha_{D}$ to be roughly -80 meV/\%, i.e. the Dirac cone is lowered in energy towards the valence band by 80 meV for each percent of tensile strain on the graphene.
Using this relation we were able to extrapolate an estimate for the zero strain band offset for all four TMDCs (see Tab.~\ref{Tab:zerostrainoffsets}). 

    \begin{table}[htb]
    \caption{Extrapolated zero-strain valence and conduction band offsets (with respect to the graphene Dirac cone) $E_V^0$ and $E_C^0$, respectively, and the absolute deformation potential $\alpha_D$, as obtained by the fit in Fig.~\ref{Fig:strainVSoffset}. Data points with excessive strain ($|\epsilon|>10\%$) or negative band offset $E_{V/C}<0$ were excluded from the fit.}\label{Tab:zerostrainoffsets}
    \begin{ruledtabular}
    \begin{tabular}{c|cccc}

				&  MoS$_2$ & WS$_2$ &MoSe$_2$ &WSe$_2$\\
				\hline
$E_V^0$[meV] &1365	    & 1027  &680	&290		\\
$E_C^0$[meV] &238		& 546	&673	&978 \\
$\alpha_{D}$ [meV/\%]  &-74          &-79       & -78      &-89

     	\end{tabular}
    \end{ruledtabular}

    \end{table}

The large magnitude of the extracted deformation potential indicates that even small strains can cause large band offset changes. 
From the perspective of perturbation theory, it is clear that the energy distance between the Dirac cone and the TMDC bands, i.e., the bands offsets, influences the proximity SOC~\citep{David2019:TwistTB2} (see App.~\ref{App:Efield}). Therefore, to obtain reliable proximity orbital and spin-orbit parameters the offsets need to be corrected.
In order to bring the Dirac cone to its apparent zero-strain level, as specified
by the obtained offsets in Tab.~\ref{Tab:zerostrainoffsets}, we apply a transverse (perpendicular to the layers) electric field to compensate the effects of strain $\epsilon$. This is done in Fig.~\ref{Fig:strainVSoffset}. The values of the compensating electric fields are listed in Tab.~\ref{Tab:param}. A field is defined as positive if it points from the TMDC layer to the graphene layer.
Figure~\ref{Fig:GlobalBS} shows the global band structure along high-symmetry lines for three selected angles, before and after the offset correction. The Dirac points are located well inside the TMDC band gaps and the proximity effects can be well discerned, see Fig.~\ref{Fig:ZoomBS}.

\section{Effective Hamiltonian}
\label{Sec:effHam}
    \begin{figure*}[htbp]
\includegraphics[width=.75\linewidth]{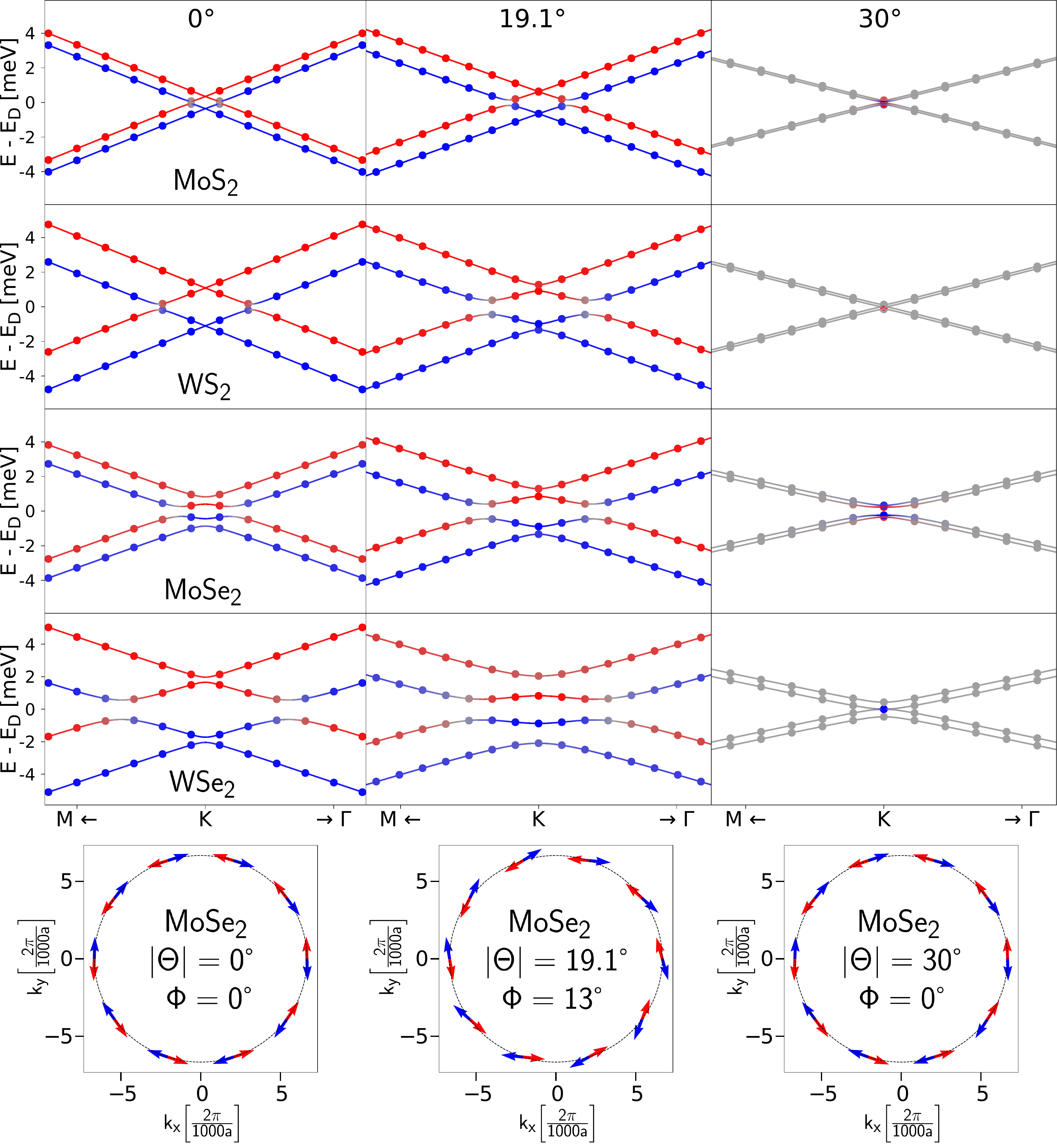} 
     \caption{Proximity Dirac band structures (after correcting band offsets for strain) at K.  We show the results for  $|\Theta|= 0\degree$, $19.1\degree$, and $30\degree$. The dots represent the DFT data while the solid line represents the fit from the model Hamiltonian in Eq. \eqref{Eq:Ham}. The spin-z expectation value is color coded from $\langle s_{z}\rangle=-0.5$ (spin down; blue) over $\langle s_{z}\rangle=0$ (unpolarized; grey) to $\langle s_{z}\rangle=0.5$ (spin up; red). The extracted parameters are listed in Tab.~\ref{Tab:param}. The last row shows the spin-x and spin-y expectation values of the spin-up (red arrows) and spin-down (blue arrows) valence band of the Dirac cone of graphene proximitized by MoSe$_2$ for the three twist angles. The k-path (dotted line) goes along a circular path around the K-point. We also present the extracted value for the Rashba angle $\Phi$ which measures the uniform deviation of the spin texture from  tangential.}\label{Fig:ZoomBS}
    \end{figure*}
To find the twist-angle dependence of the proximity induced SOC in our heterostructures, we fit the DFT band structures at the Dirac points to a model
Hamiltonian~\citep{Gmitra2015:TMDCgraphene1}. The Hamiltonian $H$ comprises the orbital part $H_{\text{orb}}$ and the spin-orbit part $H_{\text{so}}$, which is composed of the 
intrinsic spin-orbit coupling $H_{\text{so,I}}$ and the Rashba coupling $H_{\text{so,R}}$:
\begin{equation}
H(\mathbf{k})=H_{\text{orb}}(\mathbf{k})+H_{\text{so}}=H_{\text{orb}}(\mathbf{k})+
H_{\text{so,I}}+H_{\text{so,R}}.
\label{Eq:Ham}
\end{equation}
The orbital part describes the dispersion of the graphene Dirac cone linearized around the K/K'-point; accordingly, $\mathbf{k}$ is the electron wave vector measured from K/K'. It also includes a staggered potential $\Delta$, caused by the substrate's asymmetrical influence on the graphene A- and B-sublattice:
\begin{equation}
H_{\text{orb}}(\mathbf{k})=\hbar v_F (\kappa\sigma_x k_x+\sigma_y k_y)+\Delta \sigma_z.
\end{equation}
Here, $v_F$ is the Fermi velocity of the Dirac electrons and $\sigma$ are the Pauli matrices operating on the sublattice (A/B) space. The parameter $\kappa= 1$ for K and 
$\kappa=-1$ for K'.

The intrinsic spin-orbit Hamiltonian
\begin{equation}
H_{\text{so,I}}= \frac{1}{2}\Big[\lambda_{\text{I}}^A(\sigma_z+\sigma_0)+\lambda_{\text{I}}^B(\sigma_z-\sigma_0)\Big]\kappa s_z,
\end{equation}
 and the Rashba spin-orbit Hamiltonian
 \begin{equation}
H_{\text{so,R}}= -\lambda_{\text{R}} \exp(-i\Phi \frac{s_z}{2})
\Big[\kappa\sigma_x s_y-\sigma_y s_x\Big]\exp(i\Phi \frac{s_z}{2}),
\end{equation}
both additionally act on the spin space, described by the spin Pauli matrices $s$; $\lambda_{\text{I}}^A$ and $\lambda_{\text{I}}^B$ are the intrinsic spin-orbit parameters for the sublattices A and B respectively, while $\lambda_{\text{R}}$ is the Rashba SOC strength. The Rashba phase angle $\Phi$---present in $C_3$ symmetric structures~\cite{Li2019:TwistTB1, David2019:TwistTB2}---rotates the spin texture about the $z$-axis, adding a radial component to the Rashba field. For $\Theta=0\degree$~and $\Theta=30\degree$, the reflection symmetry~\citep{Li2019:TwistTB1} along the $x$- and $y$-axis respectively forces either $\Phi=0$ or $\Phi=180\degree$, therefore this parameter does not appear in the Hamiltonian of Ref.~\cite{Gmitra2015:TMDCgraphene1} valid for aligned structures. 

We choose to limit the Rashba parameter to positive values $\lambda_{\text{R}} >0$. A sign change of $\lambda_{\text{R}}$ then corresponds to an additional phase shift of $\Phi$ by a half rotation, i.e. $\Phi\rightarrow\Phi+180\degree$. To make this clear we always list $|\lambda_{\text{R}}|$ below. We also evaluated the angles $\Phi$ for a few selected supercells (see Fig.~\ref{Fig:ZoomBS} and App.~\ref{App:imagRashba}).
While the energy dispersion is not affected by $\Phi$, the radial component of the Rashba field can affect spin physics and need to be considered when interpreting experimental results on spin transport and spin-orbit torque. For example, the in-plane spin accumulation in the presence of electrical current can have a component along the current, unlike for the usual Rashba effect.

We also define 
\begin{align}
  \lambda_{\text{VZ}}=\frac{\lambda_{\text{I}}^A-\lambda_{\text{I}}^B}{2}\text{~~~and~~~}\lambda_{\text{KM}}=\frac{\lambda_{\text{I}}^A+\lambda_{\text{I}}^B}{2}  
\end{align}

as the valley-Zeeman~\cite{Gmitra2015:TMDCgraphene1, Wang2015:NC} SOC (sublattice-odd) and the Kane-Mele~\cite{Kane2005:PRL} SOC (sublattice-even) respectively. It turns out that $\lambda_{\text{KM}}$ is negligible for graphene/TMDC heterostructures~\cite{Gmitra2015:TMDCgraphene1, Wang2015:NC}. This is also true for the 
twisted heterostructures presented below, and already predicted by  tight-binding modeling~\cite{David2019:TwistTB2,Li2019:TwistTB1}.

Our DFT structures are for angles between 0\degree~and 30\degree. The results for all other twist angles can be obtained by symmetry as follows.
Twisting clockwise or counterclockwise from 0\degree~does not influence the parameters: 
\begin{align}
\lambda_{\text{VZ}}(-\theta)&=\lambda_{\text{VZ}}(\theta) \\
\lambda_{\text{R}}(-\theta)&=\lambda_{\text{R}}(\theta)\\
\Delta(-\theta)&=\Delta(\theta).
\end{align}

Additionally a twist by 60\degree~corresponds to switching the sublattices of graphene and therefore changes the sign of the sublattice-sensitive parameters:
\begin{align}
\lambda_{\text{VZ}}(\theta+60\degree)&=-\lambda_{\text{VZ}}(\theta)\\
\lambda_{\text{R}}(\theta+60\degree)&=\lambda_{\text{R}}(\theta)\\
\Delta(\theta+60\degree)&=-\Delta(\theta).
\end{align}

Using the first set of rules, one can infer the parameters' values for $\Theta\in [-30\degree,0\degree]$ from the values for $\Theta\in [0\degree,30\degree]$. Then, using the second set of rules, one can infer the values for $\Theta\in [30\degree,60\degree]$. For example, $\lambda_{\text{VZ}}(32\degree)=-\lambda_{\text{VZ}}(-28\degree)=-\lambda_{\text{VZ}}(28\degree)$. This leads to the conclusion that $\lambda_{\text{VZ}}$ changes sign at $\Theta=30\degree$ and
$\lambda_{\text{VZ}}(30\degree) = 0$.

\section{Results}
\label{Sec:Results}
\begin{figure}[ht]
     \includegraphics[width=.99\linewidth]{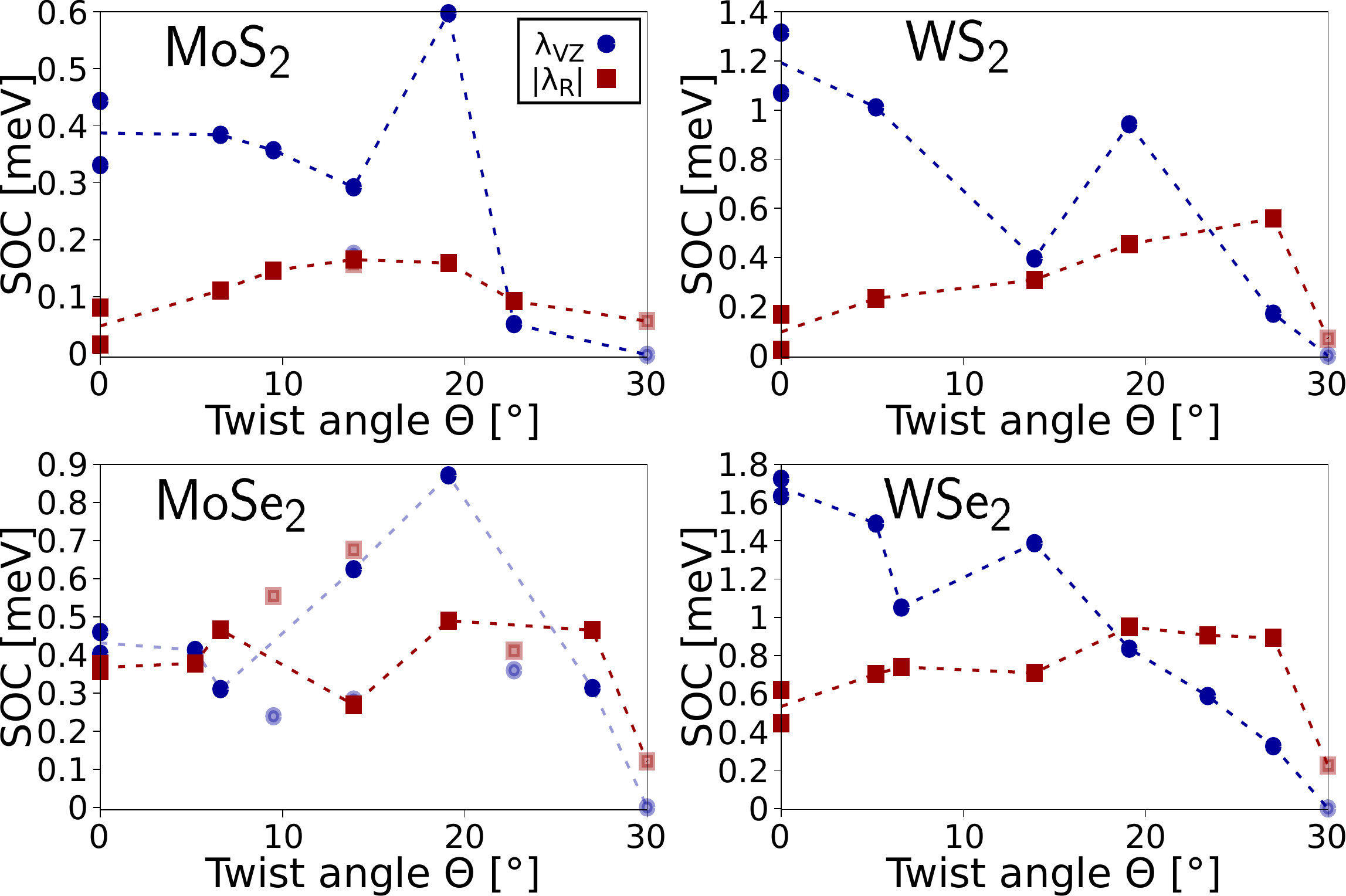}
     \caption{The E-field corrected spin-orbit parameters $\lambda_{\text{R}}$ (red squares) and $\lambda_{\text{VZ}}$ (blue dots) of the graphene Dirac cone as function of the twist angle $\Theta$ for the heterostructures with TMDCs MoS$_2$,  MoSe$_2$,  WSe$_2$, and  WS$_2$, as indicated. We de-emphasize (transparent, light red) data points from supercells with excessive built-in strain of $\epsilon>10\%$, deeming them less reliable. The dotted line is a guide for the eyes. \label{Fig:params}}
    \end{figure}

The fitted valley-Zeeman $\lambda_{\text{VZ}}$ and Rashba $\lambda_{\text{R}}$ spin-orbit couplings are collected in Tab.~\ref{Tab:param} for both uncorrected and corrected electronic band structures. The latter are plotted 
in Fig.~\ref{Fig:params}. In agreement with earlier DFT studies of aligned graphene/TMDC heterostructures~\cite{Gmitra2015:TMDCgraphene1, Gmitra2016:TMDCgraphene2, Wang2015:NC}, the proximity spin-orbit coupling of the Dirac electrons is on the meV scale. In all cases  valley-Zeeman SOC $\lambda_{\text{VZ}}$ vanishes at $\Theta=30\degree$, while the Rashba SOC $\lambda_{\text{R}}$ reaches a minimum there. We also find that $\lambda_{\text{VZ}}$ for the Mo-based TMDC heterostructures exhibit a peak at about $\theta=20\degree$, where the parameters jump to about twice the value at 0$\degree$. This pronounced peak is preceded by a slight dip at around $\Theta$ between 10$\degree$ and 15$\degree$. Similarly, for the heterostructures based on WSe$_2$ and WS$_2$ there appears to be a peak in the valley-Zeeman coupling below 20$\degree$, although its magnitude is less that the value at 0$\degree$. Except for twist angles close to 30$\degree$, the Rashba parameters are generally smaller in magnitude than $\lambda_{\text{VZ}}$.

Tight-binding models~\cite{Li2019:TwistTB1,David2019:TwistTB2} predicted a peak structure with a global maximum in the region of 15$\degree$-20$\degree$. We can confirm this picture for Mo-based but not for W-based heterostructures. The two tight-binding models differ in their predictions for the sign of $\lambda_{\text{VZ}}$: Ref.~\cite{Li2019:TwistTB1} predicts positive $\lambda_{\text{VZ}}$, while Ref.~\cite{David2019:TwistTB2} predicts a sign change at around 10$\degree$. As seen in Fig.~\ref{Fig:params}, the extracted valley-Zeeman coupling parameters do not change sign, in agreement with Ref.~\cite{Li2019:TwistTB1}.

To check our approach, for 0\degree~---and for MoS$_2$ and MoSe$_2$ additionally for 13.9\degree---we plot in Fig.~\ref{Fig:params} two data points corresponding to two different supercells with the same twist angle but different strain. Such data pairs indicate how reliable the DFT approach based on strained supercells is. Ideally, the extracted parameters would agree for the given twist angle after correcting for strain by transverse electric field. This is indeed the case 
at 0\degree~ where data pairs for TMDCs agree rather
well. However, for the 13.9\degree~data pairs the parameters differ significantly. At this angle one of the heterostructures in the pairs have strain above 10\%. We believe that such a strain is already too large for obtaining quantitatively reliable proximity band structures even after correcting by the electric field. We exclude such data from considering the trends seen in Fig.~\ref{Fig:params}. The reason for the sensitivity to strain is that the size of the Brillouin zone
of strained graphene varies with $\epsilon$. Following the interlayer coupling picture of Refs.~\citep{Koshino2015:TwistTBBasic, David2019:TwistTB2}, the graphene K-point thus couples to somewhat different regions of the TMDC Brillouin zone. This momentum-specific coupling (hybridization of the corresponding graphene and TMDC orbitals) is responsible for the proximity effect. 

We note that the differences in our reported data in 
Tab.~\ref{Tab:param} and previous ones~\cite{Gmitra2016:TMDCgraphene2} at 0$\degree$ are due to the unrelaxed atomic structures used here. The main effect of atomic relaxation is to introduce rippling of the graphene layer which enhances especially the staggered potential $\Delta$ and the Kane-Mele SOC  $\lambda_{\text{KM}}$ parameter. In our calculations both $\Delta$ and $\lambda_{\text{KM}}$ are rather suppressed. We have studied the atomic relaxation effects for selected heterostructures with 0 and 19.1$\degree$ twist angles; the results are presented in App.~\ref{App:relax}).

Finally, let us compare the proximity SOC for uncorrected and corrected band structures. The former
is plotted in Fig.~\ref{Fig:paramsnoE}, the latter in Fig.~\ref{Fig:params} which we discussed above. We see that the
while the overall magnitudes of $\lambda_{\text{VZ}}$ and $\lambda_{\text{R}}$ are similar for both cases, the correction by electric field has an influence on the overall twist-angle dependence. The main effect of the correction is shifting the band offsets which modify the interlayer coupling and charge transfer between the layers. Both have an effect on the valley-Zeeman and Rashba couplings. In App.~\ref{App:Efield} we further investigate the influence of the electric field on the proximity spin-orbit parameters for a 19.1$\degree$ graphene/MoS$_2$ heterostructure, demonstrating a rather strong decrease of the magnitude of $\lambda_{\text{VZ}}$ upon increasing of E-field and accompanying increase of the band offset $E_C$, see Fig.~\ref{Fig:Efield}.

\section{Summary}

We have performed systematic investigations of the twist-angle dependence of the proximity spin-orbit coupling induced by four TMDC monolayers MoS$_2$, MoSe$_2$, WS$_2$, and WSe$_2$ in graphene. For each graphene/TMDC bilayer we have studied several twist angles, being computationally limited by the supercell sizes. We have identified strain, which is necessary to build commensurate supercells of reasonable sizes, as the key factor influencing the reliability of the extracted SOC parameters. To correct for strain, we apply a transverse (to the layers) electric field which compensates the changes of the band offsets caused by strain.

Extracting the SOC parameters by fitting the corrected DFT band structures to an already established model Hamiltonian, we present as the main result the magnitudes of the valley-Zeeman and Rashba couplings of the Dirac electrons as functions of the twist angles from 0$\degree$ to 30$\degree$; symmetry relations can be used to deduce the parameters for other angles. A qualitatively new effect of twisting is the emergence of the Rashba phase angle $\Phi$ which rotates the spin texture away from tangential; $\Phi$ vanishes at 0$\degree$ and 30$\degree$. There appears to be no discernible trend for $\Phi$ as a function of the twist angle, but we show that the phase angle depends strongly on the electric field. While not affecting the energy dispersion, nonzero $\Phi$ adds a radial component to the typical tangential Rashba texture and should be considered when interpreting spin transport experiments in twisted heterostructures. In particular, the spin accumulation due to electric current need not be perpendicular to the current. Unfortunately, the phase angles appear too sensitive to the band offsets and transverse electric fields, preventing us to make reliable quantitative predictions on their magnitudes. It is natural to assume that each twisted heterostructure is in this sense unique, exhibiting $\Phi$ based on the twist angle, unintended doping, interlayer distance, and applied gate. An experimental observation of $\Phi$ and its tunability 
is currently outstanding.

As required by symmetry, the valley-Zeeman SOC vanishes at 30$\degree$. This clear case should serve as a useful tool for experiments relying on valley-Zeeman coupling. For other twist angles, we can conclude that the coupling does not change sign between 0$\degree$ and 30$\degree$, 
and that $\lambda_{\text{VZ}}$ for Mo-based TMDCs
has a (global) maximum at roughly 20\degree. For
W-based heterostructures the maximum coupling appears to be at 0$\degree$. The Rashba coupling is, in general, weaker than the valley-Zeeman coupling, except close to 30$\degree$ where the latter vanishes. 

We have also studied the effects of structural relaxation on the proximity effects. We found that rippling of graphene is the main factor significantly enhancing the staggered potential $\Delta$ and the Kane-Mele coupling $\lambda_\text{KM}$ (which is otherwise negligible for unrelaxed structures), but does not affect strongly the valley-Zeeman and Rashba couplings. Relative lateral shifts of the graphene and TMDC layers too do not have a significant effect on the proximity spin-orbit parameters.

\acknowledgments This work was funded by the International Doctorate~Program Topological~Insulators of the Elite~Network of Bavaria, the Deutsche Forschungsgemeinschaft (DFG, German Research Foundation) SFB 1277 (Project-ID 314695032), SPP 2244 (project no. 443416183), and by the European Union Horizon 2020 Research and Innovation Program under contract number 881603 (Graphene Flagship). M.G. acknowledges VEGA 1/0105/20.

\appendix
 \begin{table*}[htb]
   \caption{Parameters extracted from the band structure calculations. For all four TMDCs and all used angles, we list the band offset $E_V$ ($E_C$) of the Dirac cone with respect to the valence band (conduction band) and the extracted model Hamiltonian (see Eq. \eqref{Eq:Ham}) parameters (staggered potential $\Delta$, intrinsic SOC for site A and B $\lambda_{\text{I}}^A$ and $\lambda_{\text{I}}^B$ and Rashba SOC $\lambda_{\text{R}}$). We denote the offsets and parameters after correction with the electric field with a bar, for example $\bar{\lambda}_I^A$. The electric field is defined as positive, if it points from the TMDC layer to the graphene layer.}
    \label{Tab:param} 
    \begin{ruledtabular}
    \begin{tabular}{cc|cccccc|c|cccccc}

$|\theta|[\degree]$&$\epsilon$&$\Delta$ & $\lambda_{\text{I}}^A$ & $\lambda_{\text{I}}^B$ & $|\lambda_{\text{R}}|$& $E_V$& $E_C$& E-field& $\bar{\Delta} $&$\bar{\lambda}_I^A $ & $\bar{\lambda}_I^B $ & $|\bar{\lambda}_R|$& $\bar{E}_V$& $\bar{E}_C$ \\

&[\%]&[meV] & [meV] & [meV] & [meV]& [eV]&[eV]& [$\text{V}/\text{nm}$]& [meV] &[meV]  & [meV] & [meV]& [eV]& [eV] \\
\hline
MoS$_2$&&&&&&&&\\
\hline
0 & -2.9& 0.014 & 0.346 & -0.346 & 0.085 &1.598  &0.004  		&1.499		&0.009 &0.331 &-0.331 &0.081 & 1.341 &0.262  \\
0 & 3.58& -0.005 & 0.402 & -0.408 & 0.043&1.066 &0.533                                      &-1.748 &  0.130 & 0.445 & -0.442 & 0.016&1.358&0.236\\
5.2 &-4.99 &0.022 &0.455 &-0.457 & 0.165 &1.773 &-0.169 	            & -	    & - & - & - & - & -& - \\
6.6 &2.89 &0.054 &0.341 &-0.341 &0.104 &1.118 &0.482 			& -1.301		& 0.032 & 0.384 & -0.384 & 0.111 & 1.334 &0.264 \\
9.5 &7.1 &0.023 &0.225 &-0.225 &0.139 &0.838 &0.761			&-2.962		& 0.026 & 0.356 & -0.358 & 0.146 & 1.322 &0.262  \\
13.9 &7.73  &0.017 &0.119 &-0.120 &0.168 &0.809  &0.791		&-3.183		& 0.010 & 0.291 & -0.293 & 0.165 & 1.331 &0.258  \\
13.9 & 16.7 &0.037 & -0.270 & 0.270 & 0.265   &0.377&1.230                         &-6.154 &    -0.327 & 0.176 & -0.174 & 0.157  &1.381&0.256 \\
13.9 &-6.64 & 0.124 & 8.707 & -8.754 & 1.251&1.828 & -0.222                                   & -		    & - & - & - & - & -& -\\
19.1 &-2.13 &0.023 &1.223 &-1.230 &0.241 &1.532   &0.075		&1.107 		& 0.031 & 0.593 & -0.602 & 0.159 & 1.336 &0.258 \\
22.7 &7.1	 &-0.005 &-0.040 &0.040 &0.047 &0.859 &0.745		 &-2.564		& -0.004 & 0.052 & -0.052 & 0.092 & 1.242 &0.315 \\
27  &-4.99 &-0.004 &0.459 &-0.458 &0.531 &1.773 &-0.168  		& -		& - & - & - & - & -& - \\
30 &12.13 &0.000 & 0.000 &0.000 &0.075 &0.596 &1.009 		&-6.582		& 0.059 & 0.002 & 0.002 & 0.057 &1.359 &0.235 \\
30 & -10.3 &0.000 & -0.062 & -0.062 & 1.774 &1.868 &-0.418	& -		& - & - & - & - & -& -\\
\hline
WS$_2$&&&&&&&&\\
\hline
0 &	-3.05 &0.021 & 1.094 & -1.095 & 0.154  &1.248 &0.325       &0.951     & 0.014 & 1.068 & -1.072 & 0.168 &1.082 &0.492\\
0 &3.41 & -0.097 & 1.201 & -1.193 & 0.077& 0.717 &0.853                 & -1.969  &  -0.047 & 1.32 & -1.312 & 0.024&1.053&0.514\\
5.2 &-5.14 & 0.000 & 1.148 & -1.152 & 0.287 &1.459 &0.114        &2.234     & 0.008 & 1.011 & -1.012 & 0.233 &1.056 &0.518  \\
13.9 & 7.56    & 0.866 & -0.076 & 0.068 & 0.386      &0.463&1.110                           & -3.416     &  -0.263 & 0.394 & -0.397 & 0.309   &1.035&0.531     \\
19.1 &-2.28 & 0.004 & 1.305 & -1.316 & 0.526 &1.185 &0.389    &0.662       & 0.005 & 0.938 & -0.947 & 0.454 &1.067 &0.507 \\
27 &-5.14 & -0.052 & 0.608 & -0.606 & 0.959 &1.459 & 0.116      &2.195     & 0.010 & 0.171 & -0.170 & 0.559  &1.073 &0.503 \\
30 &11.95 & 0.000 & 0.004 & 0.004 &0.214 &0.237 &1.327      &-4.973 & -0.001 & 0.004 & 0.004 & 0.069 &1.067 &0.504\\

\hline

MoSe$_2$&&&&&&&&\\
\hline
0 &1.19 & 0.034 & 0.418 & -0.416 & 0.425 & 0.535 &0.817		&-0.863		& 0.017 & 0.404 & -0.402 & 0.370	&0.673 &0.679\\
0 & 7.93& -0.625 & 0.282 & -0.271 & 0.717&0.098 &1.252                                  & -3.754 &  0.101 & 0.463 & -0.457 & 0.358&0.667&0.678\\
5.2 &-1 & 0.024 & 0.412 & -0.412 & 0.368 &0.708 &0.645 		&0.209		& 0.016 & 0.413 & -0.413 & 0.378&0.674 &0.679	\\
6.6 &7.22 &0.069 &-0.023 &0.030 &0.834 & 0.142 & 1.209 		&-3.453		& -0.006 & 0.311 & -0.310 & 0.466 	&0.672 &0.676	\\

9.5 &11.6 & 0.038 & -1.099 & 1.061 & 1.380 &-0.051 &1.386 		&-5.101		& 0.038 & 0.239 & -0.239 & 0.555 &0.687 &0.657	 \\
10.9 &-11.67&0.244 &4.906 &-4.632 &1.011 & 1.719 & -0.363		&-&-&-&-&-&-&-\\
13.9 &12.26 & 0.029 & -0.994 & 0.923 & 1.621 & -0.069 & 1.433 		&-5.150		& -0.051 & 0.285 & -0.281 & 0.676 &0.675 &0.672\\
13.9 & -2.71 & 0.024 & 0.694 & -0.696 & 0.227  &0.850& 0.503                                   &1.080   &  -0.016 & 0.624 & -0.626 & 0.269   &0.676&0.677\\
19.1 &1.99 & -0.015 & 0.873 & -0.873 & 0.573 & 0.496 & 0.856 		 &-1.048		& 0.000 & 0.871 & -0.872 & 0.490&0.655 &0.678\\
22.7 &11.6 &-0.013 &-0.028 &0.029 &0.881 & -0.026 & 1.370 		&-4.958 & 0.033 & 0.360 & -0.360 & 0.411  &0.696   &0.655	 \\
23.4 &-7.14& -0.002 & -0.209 & 0.213 & 0.253& 1.284 & 0.071 	&-&-&-&-&-&-&-\\
27 &-1	 & -0.012 & 0.306 & -0.305 & 0.446 &0.721 & 0.633  	& 0.288		& -0.012 & 0.313 & -0.313 & 0.465&0.677 &0.677	\\
30 &16.84 &0.000 &0.005 &0.005 &0.509 & -0.178 & 1.544 		&-6.515		& -0.236 & 0.003 & 0.003 & 0.121 	&0.675 &0.655 \\
30 &-6.53& 0.000 & -0.004 & -0.004 & 0.213& 1.224 & 0.131 &-&-&-&-&-&-&-\\

\hline
WSe$_2$\\
\hline
0 &1.19 & 0.042 & 1.651 & -1.647 & 0.671 &0.186 &1.081 		&-0.636			& 0.056 & 1.635 & -1.633 & 0.621 &0.289 &0.978 \\
0 &7.93& 0.219 & 1.537 & -1.519 & 0.671&-0.018 &1.282                                 &-3.445           &  0.084 & 1.728 & -1.722 & 0.445&0.286&0.977\\
5.2 &-1 & 0.034 & 1.495 & -1.494 & 0.675 &0.360 &0.907 		 &0.449		& 0.033 & 1.491 & -1.490 & 0.703 &0.291 &0.977\\
6.6 &7.22 &0.090 & -0.161 &0.169 &1.219 & -0.164 & 1.429  	 &-3.137		& 0.073 & 1.052 & -1.050 & 0.740  &0.285 &0.979\\
10.9 & -11.67 &0.014 & 3.185 & -3.211 & 1.051 &1.417 &-0.147 				&- &- &- &- &- &-&-\\
13.9 & -2.71   & 0.023 & 1.612 &-1.616 & 0.642         &0.504 & 0.764            &  1.327     &  -0.022 & 1.387 & -1.388 & 0.709    &0.290 &0.978\\
19.1 &1.99 & 0.009 & 0.640 & -0.631 & 1.051 &0.138 &1.123 		&-0.920		& 0.014 & 0.836 & -0.835 & 0.950 &0.294 &0.974 \\
23.4 &-7.14 & -0.004 & 0.641 & -0.644 & 0.911 & 0.943 &0.326 	&3.085			&0.000 &0.588 &-0.588 &0.906 &0.427 & 0.783 \\
27 &-1 & 0.000 & 0.347 & -0.343 & 0.868 & 0.375 &0.894 		&0.555			& 0.000 & 0.328 & -0.323 & 0.892  &0.294 &0.975 \\
30 &16.84 & 0.001 & 0.018 & 0.019 & 0.198 &-0.349 &1.633 	&-6.065			&0.003 &0.009 &0.009 &0.225 &0.313 &0.901 \\

\hline

    \end{tabular}
    \end{ruledtabular}
    \end{table*}

\section{Computational details}
\label{App:A}

	All electronic structure calculations
	are performed by density functional theory (DFT)~\citep{Hohenberg1964:PRB} 
	with {\tt Quantum ESPRESSO}~\citep{Giannozzi2009:JPCM}.
	Self-consistent calculations are carried out with a $k$-point sampling of $n_k\times n_k\times 1$. The number $n_k$ is listed in Table~\ref{Tab:ks} for all twist angles.
	We use energy cutoffs listed in Tab.~\ref{Tab:ks} for the scalar relativistic pseudopotential
	with the projector augmented wave method~\citep{Kresse1999:PRB} with the 
	Perdew-Burke-Ernzerhof exchange correlation functional~\citep{Perdew1996:PRL}. Graphene's $d$-orbitals are not included in the calculations. We used 
    Van der Waals corrections~\citep{Grimme2006:JCC,Grimme2010:JCP,Barone2009:JCC}.
	
 The electric fields are implemented in the DFT calculations using a sawtooth potential in $z$-direction within the quasi 2D unit cell. The electric potential increases linearly in the area of the heterostructure and then falls rapidly in the vacuum. 
    \begin{table}[htb]
    \caption{Computational details: Used charge density cutoff energy $E_\rho$, wave function kinetic energy cutoff $E_{\text{wfc}}$ and $k$-grid density (we used a $n_k\times n_k$ grid) for the calculations listed in Tab.~\ref{Tab:param}. E-field calculations are listed with a tilde over the material. The supercells are in the same order as in Tab.~\ref{Tab:Structures}.}\label{Tab:ks}
    \begin{ruledtabular}
    \begin{tabular}{c|cccccccc}

				&  MoS$_2$ 	&$\widetilde{\text{MoS}}_2$	&WS$_2$&$\widetilde{\text{WS}}_2$ &MoSe$_2$&$\widetilde{\text{MoSe}}_2$ 	&WSe$_2$&$\widetilde{\text{WSe}}_2$ \\
				\hline
$E_\rho$[Ry] 			& 55 	& 55 &70	& 70	&	60	&	60	&	65	&65	 \\
$E_{\text{wfc}}$[Ry]	& 350	& 350 & 500& 500	&	350	&350	&   550&   550		\\
\hline
$n_k$\\
\hline

0\degree	&21	&21 &15	&15	&21	&15	&21	&15		\\
0\degree    &3  &9  &3  &9  &3  &9  &3  &9   \\
5.2\degree	&21	&-	&21	&15 &21	&15	&21	&21		\\
6.6\degree	&21	&21	&-	&-  &18	&12	&18	&12		\\
9.5\degree	&21	&18	&-	&-  &18	&18	&-	&-		\\
10.9\degree	&-	&-	&-	&-  &21	&-	&21	&-		\\
13.9\degree	&21	&21	&3	&9  &21	&15	&-	&-		\\
13.9\degree &9  &15 &-  &-  &-  &-  &-  &-   \\
13.9\degree &21 &-  &-  &-  &6  &12 &3  &3   \\
19.1\degree	&30	&30	&30	&30&21	&21	&21	&21		\\
22.7\degree	&21	&18	&-	&-&18	&12	&-	&-		\\
23.4\degree	&-	&-	&-	&-&18	&-	&21	&12		\\
27\degree	&21	&-		&15	&15&21	&15	&21	&21	\\
30\degree	&39	&39	&21	&36&21	&21	&21	&21		\\
30\degree	&21	&-	&-	&-&6	&-	&-	&-		\\
     	\end{tabular}
    \end{ruledtabular}

    \end{table}

\section{Effect of lateral shift between the layers}

For incommensurate heterostructures consisting of two materials whose lattice constants are no integer multiples of each other, the lateral shifting degree of freedom does not exist. If the sample can be assumed to be infinite in $x$- and $y$-direction, every shifting configuration exists somewhere on the sample. The physical properties of the configurations will then average out, when considering the properties of the whole material.

However, the structures we consider in our DFT calculations are commensurate to be
computationally viable and the lattice constants are forced by strain to be compatible. Therefore, different lateral relative shifts might ensue different physical properties including different proximity induced SOC. Naturally, this effect is less relevant for larger supercells, for which the averaging over the different shifts can occur within the supercell. 

\begin{figure}[htb]
     \includegraphics[width=.99\columnwidth]{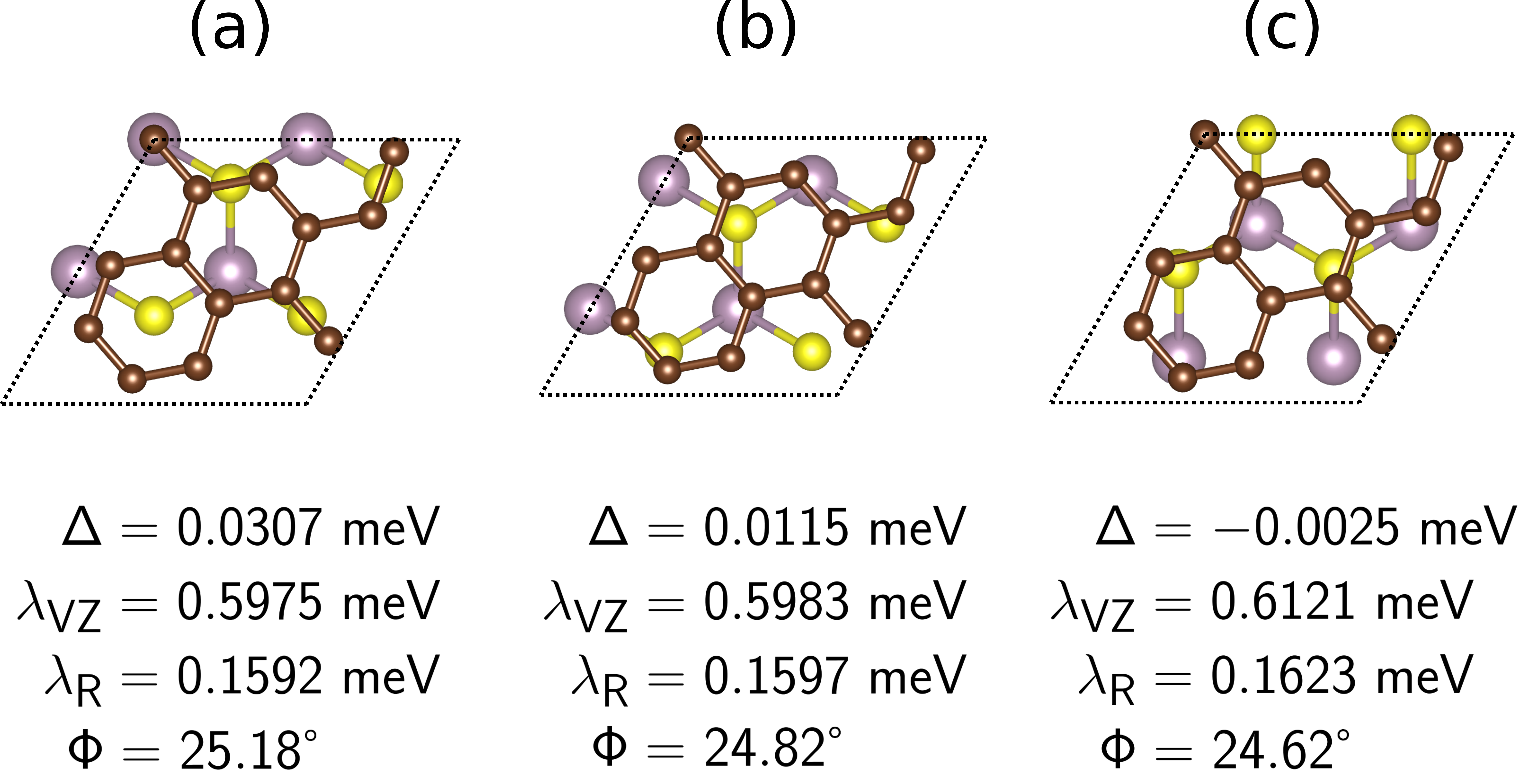}
     \caption{Three different options for lateral shifting for the MoS$_2$ 19.1\degree~supercell and their (electric field corrected) model fit parameters.}\label{Fig:shift}
    \end{figure}

In order to estimate the effects of lateral shifts we look at different configurations for one of our supercells: the 19.1\degree~supercell of MoS$_2$. As one can see in Fig.~\ref{Fig:shift}, all three supercells have very similar spin-orbit parameters and only the staggered potential $\Delta$ changes. While we cannot make a sweeping conclusion based on one twist angle and a few lateral shifts (and computational power limits one's capabilities here), these results indicated that the sizes of our supercells are already sufficient to yield quantitatively reasonable results for the magnitudes of the proximity valley-Zeeman and Rashba spin-orbit couplings. 

\label{App:shift}

\section{Effects of structural relaxation}

All the results in the main text were obtained without prior relaxation of the atomic structure. To study how atomic relaxation affects proximity SOC, we performed calculations on selected supercells which were relaxed in different ways. The results are gathered in Tab.~\ref{Tab:relax}. Our main finding is that, while the valley-Zeeman SOC $\lambda_{\text{VZ}}$ is largely unaffected, the staggered potential $\Delta$ and the Kane-Mele SOC $\lambda_{\text{KM}}$, which are both very small for the unrelaxed structures, are greatly enhanced by the relaxation. This can be traced to the rippling of graphene, which results from the relaxation. We checked this by flattening the relaxed graphene, which exhibited again very small $\Delta$ and  $\lambda_{\text{KM}}$. Our calculations indicate that even very small ripplings can induce a staggered potential $\Delta$ on the meV scale. Since the unrelaxed structures show almost zero $\Delta$ or $\lambda_{\text{KM}}$ (on the order of few tens of $\mu$eV), the rippling can be considered the sole cause for the sublattice-asymmetric parameters $\Delta$ and $\lambda_{\text{KM}}$ to be non-zero in graphene/TMDC heterostructures..

 \label{App:relax}
    \begin{table*}[htb]
    \caption{Relaxation and its effect on the Dirac cone. We examine the 0\degree~and the 19.1\degree~supercells of MoS$_2$ and the 19.1\degree~supercell of WSe$_2$. We additionally consider the 19.1\degree~MoS$_2$ supercell, with strained TMDC instead of strained graphene (marked with *). For each of those cases we distinguish between calculations without prior relaxation ('fixed'), with prior relaxation ('relaxed') and with prior relaxation, but with flattened Graphene ('flat Gr'). The latter was achieved by setting the C atom's $z$-coordinate to an average value after the relaxation. For each case we list the important structural parameters (rippling, interlayer distance $d$ and chalcogen-chalcogen distance $d_{\text{XX}}$) and the extracted parameters from the model fit ($\Delta$, $\lambda_{A}$,  $\lambda_{B}$ and $\lambda_{\text{R}}$). All calculations were performed without the electric field corrections. \label{Tab:relax}}
    \begin{ruledtabular}
    \begin{tabular}{l|lll|llll}
       &rippling[m\AA]     &d[\AA]    &$d_{\text{XX}}[$\AA] &  $\Delta$[meV]      &$\lambda_{A}$[meV]        &$\lambda_{B}$[meV]&$\lambda_{\text{R}}$[meV]\\
\hline
MoS-0\degree  &&&&&& \\
\hline
fixed           &0.00               &3.30     &3.138  		 		&0.014           &0.35            &-0.35    &0.085\\
relaxed         &368.20             &3.11  &3.128   				&2.985           &0.33            &-0.14    &0.206\\
flat Gr         &0.00               &3.33  &3.128   				&0.003           &0.30            &-0.30    &0.074      \\

\hline

MoS-19.1\degree  &&&&&&    \\
\hline
fixed               &0.00         &3.30     &3.138  			&0.023           &1.22            &-1.23    & 0.241\\
relaxed             &23.63        &3.31  &3.126  		    &-1.016          &1.38            &-0.96        &0.252\\
flat Gr             &0.00         &3.31  &3.126  		    &-0.175          &1.15            &-1.10        & 0.229\\

\hline
MoS-19.1\degree*  &&&&&&    \\
\hline
fix             &0.00                  &3.30     &3.138 				&0.018                  &0.86   &-0.87  & 0.223\\
relaxed         &8.64                    &3.37 &3.079  				     &-0.294                &0.59    &-0.53 &0.163\\
flat Gr         &0.00                  &3.31  &3.312  			        &-0.082                 &0.65  &-0.64   &0.198\\

\hline

WSe-19.1   &&&&&&   \\
\hline
fix             &0.00                  &3.30     &3.364                     &0.009          &0.64   &-0.63  &1.051\\
relaxed         &6.56            &3.37  &3.347                    &0.295          &0.60            &-0.51   & 0.763\\

    \end{tabular}
    \end{ruledtabular}
    \end{table*}

\section{Rotation of the Rashba term: Fit results}

\begin{figure}[htb]
    \centering
    \includegraphics[width=.99\columnwidth]{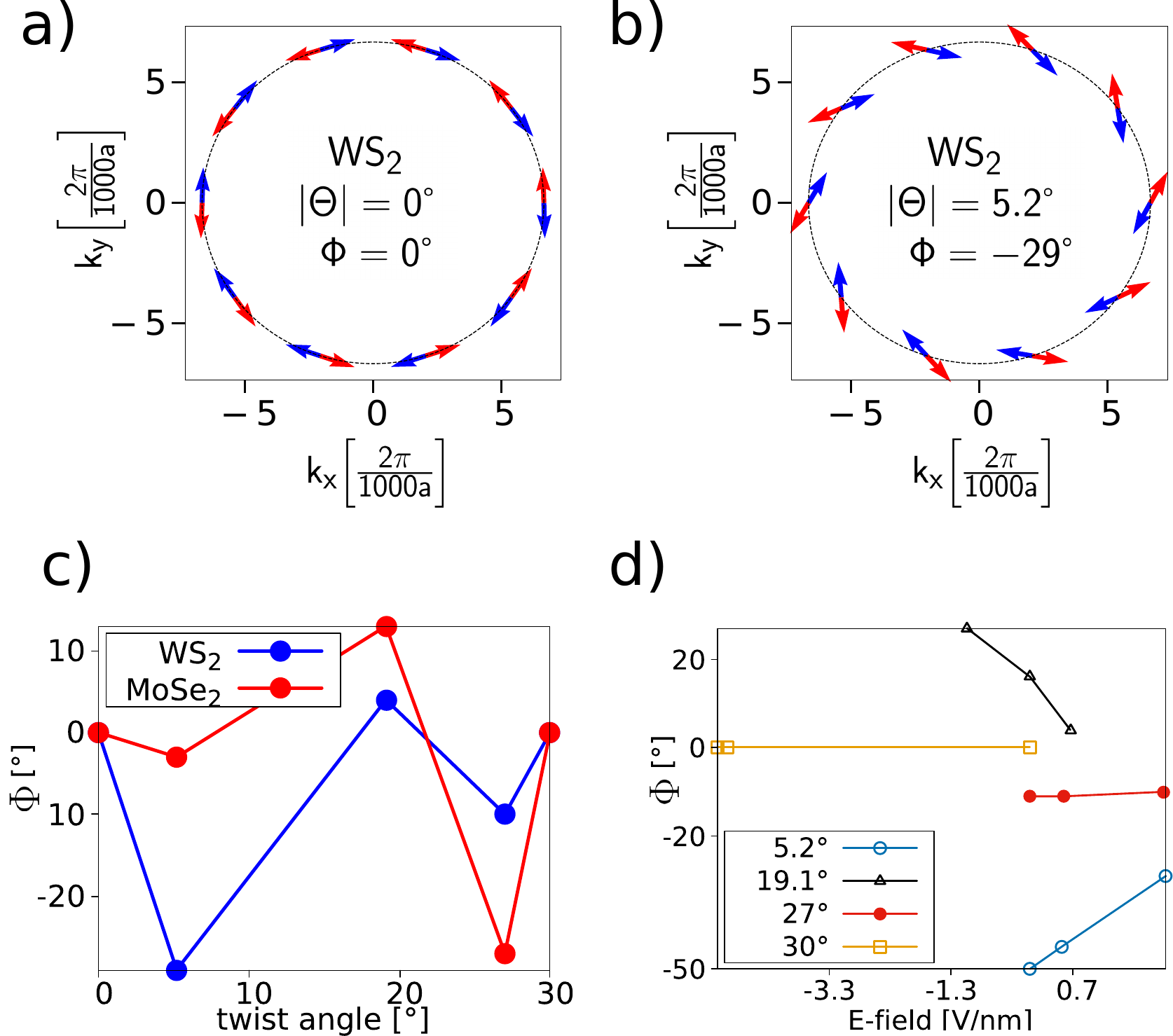}
    \caption{The parameter $\Phi$, as obtained by fitting the spin-x and spin-y expectation values (for better readability, $\Phi$ is always plotted modulo 180\degree): (a) Spin-x and spin-y expectation values of the spin-up (red arrows) and spin-down (blue arrows) valence band of the graphene Dirac cone
    proximitized by WS$_2$ with a twist angle of 0\degree. The k-path (dotted line) goes along a circular path around the K-point. Here the reflection symmetry along the $x$-axis demands $\Phi=0\degree$ or $\Phi=180\degree$. (b) shows the same as (a), but for a 5.2\degree~supercell. (c) The angle $\Phi$ extracted from the fits for several heterostructures (see Tab.~\ref{Tab:imagRashba}) plotted against the twist angle $|\theta|$. All results shown here are done with the electric field corrections. For 0\degree~and 30\degree, we extract $\Phi=0$ and $\Phi=180\degree$ as expected. (d) Electric field dependence of $\Phi$ for four WS$_2$/graphene heterostructures. Note that in c) and d) for the 5.2$^\circ$ and the 27$^\circ$ cases, we used supercells with $\theta < 0$. Since the Rashba phase angle also changes sign, when changing the sign of the twist angle, the reported phase angles are misleading. We expect that for the given twist angles with a uniform sign, the Rashba phase angles will also have uniform sign.}
    \label{Fig:imagRashba}
\end{figure}

As mentioned in the main text, to accurately describe the spin-x and spin-y expectation values of graphene in twisted heterostructures without a mirror reflection symmetry~\cite{David2019:TwistTB2, Li2019:TwistTB1}, one needs to implement a rotation of the Rashba term $H_{\text{SOC,R}}$ around the $z$-axis. We show the parameter $\Phi$, as extracted from our fits, for selected heterostructures in Tab.~\ref{Tab:imagRashba} and Fig.~\ref{Fig:imagRashba} (c). For $\Theta=0\degree$~and $\Theta=30\degree$~the reflection symmetry of the system demands $\Phi=0\degree$ or $\Phi=180\degree$. Our DFT results confirm this. 

However, there appears to be no distinct dependence of $\Phi$ on the twist angle. What we can establish is that $\Phi$ 
depends strongly on the applied transverse electric field, see
Fig.~\ref{Fig:imagRashba},
but is rather insensitive to the lateral shift of the graphene and TMDC layers at a given twist angle, see Fig.~\ref{Fig:shift}.

\begin{table}[htb]
    \centering
    \begin{tabular}{c|c|c|c}
         Material&twist angle $\theta$ [\degree] &E-field [V/nm] &$\Phi$[\degree]  \\ \hline
         WS$_2$& 0          &0                      &0                  \\
         WS$_2$&0           &0.951                  &0                  \\
         WS$_2$&-5.2         &0                      &-50                  \\
         WS$_2$&-5.2         &0.520                  &-45                  \\
         WS$_2$&-5.2         &2.234                  &-29                  \\
         WS$_2$&19.1        &0                      &16                  \\
         WS$_2$&19.1        &-1.039                 &27                \\
         WS$_2$&19.1        &0.662                  &4                  \\
         WS$_2$&-27          &0                      &-11                  \\
         WS$_2$&-27          &0.553                  &-11                  \\
         WS$_2$&-27          &2.195                  &-10                  \\
         WS$_2$&30          &0                      &180                  \\
         WS$_2$&30          &-5.142                 &180                  \\
         WS$_2$&30          &-4.973                 &180                  \\
         MoSe$_2$&0         & -0.863                &0                   \\
         MoSe$_2$&-5.2       &  0.209                &-3                   \\
         MoSe$_2$&19.1      & -1.048                &13                   \\
         MoSe$_2$&-27        &  0.288                &-27                   \\         
        MoSe$_2$&30         &  -6.515               &0                   \\
        MoS$_2$&19.1        & 0                     &41                   \\        
         MoS$_2$&19.1       & 1.107                 &25                   \\         
    \end{tabular}
    \caption{Extracted parameter $\Phi$ for selected twisted graphene/TMDC heterostructures.}
    \label{Tab:imagRashba}
\end{table}
\label{App:imagRashba}

\section{Effect of electric field}

\label{App:Efield}

\begin{figure}[htb]
     \includegraphics[width=.99\columnwidth]{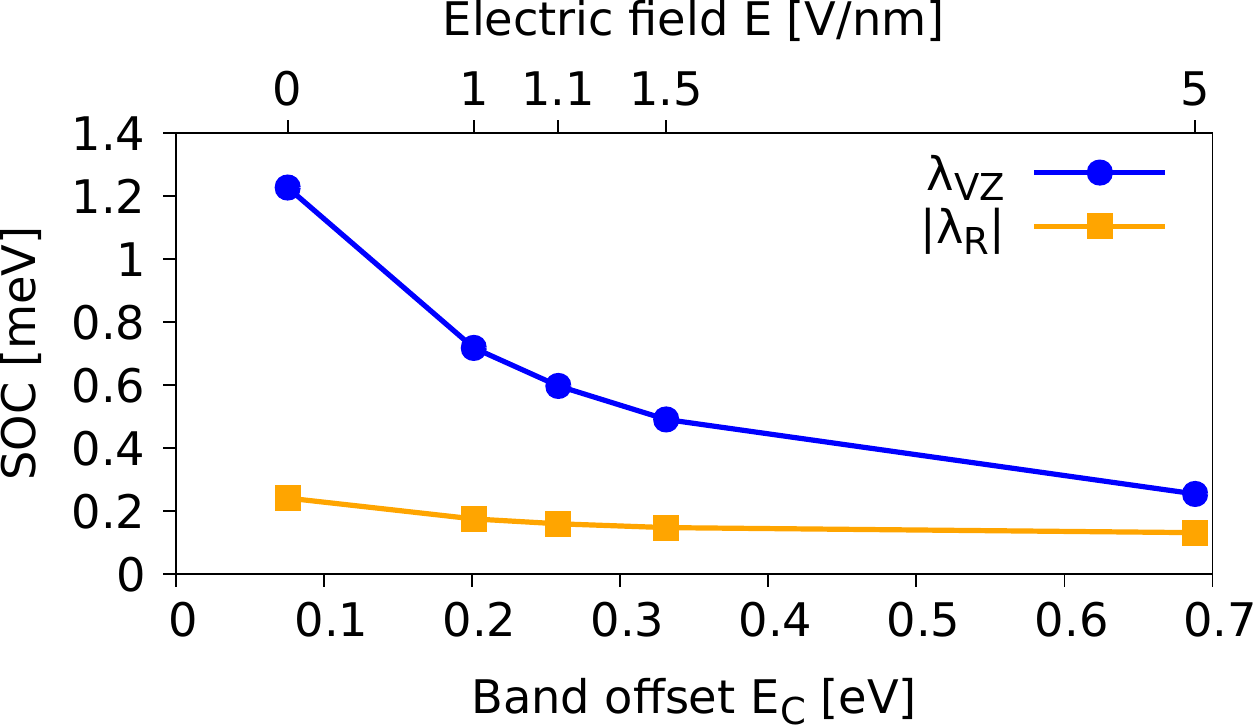}
     \caption{The effect of the correcting electric field on the spin-orbit parameters for the MoS$_2$ 19.1\degree~supercell. The electric field increases the conduction band offset $E_C$  and therefore reduces the proximity induced SOC.}\label{Fig:Efield}
    \end{figure}

\begin{figure}[htb]
     \includegraphics[width=.99\linewidth]{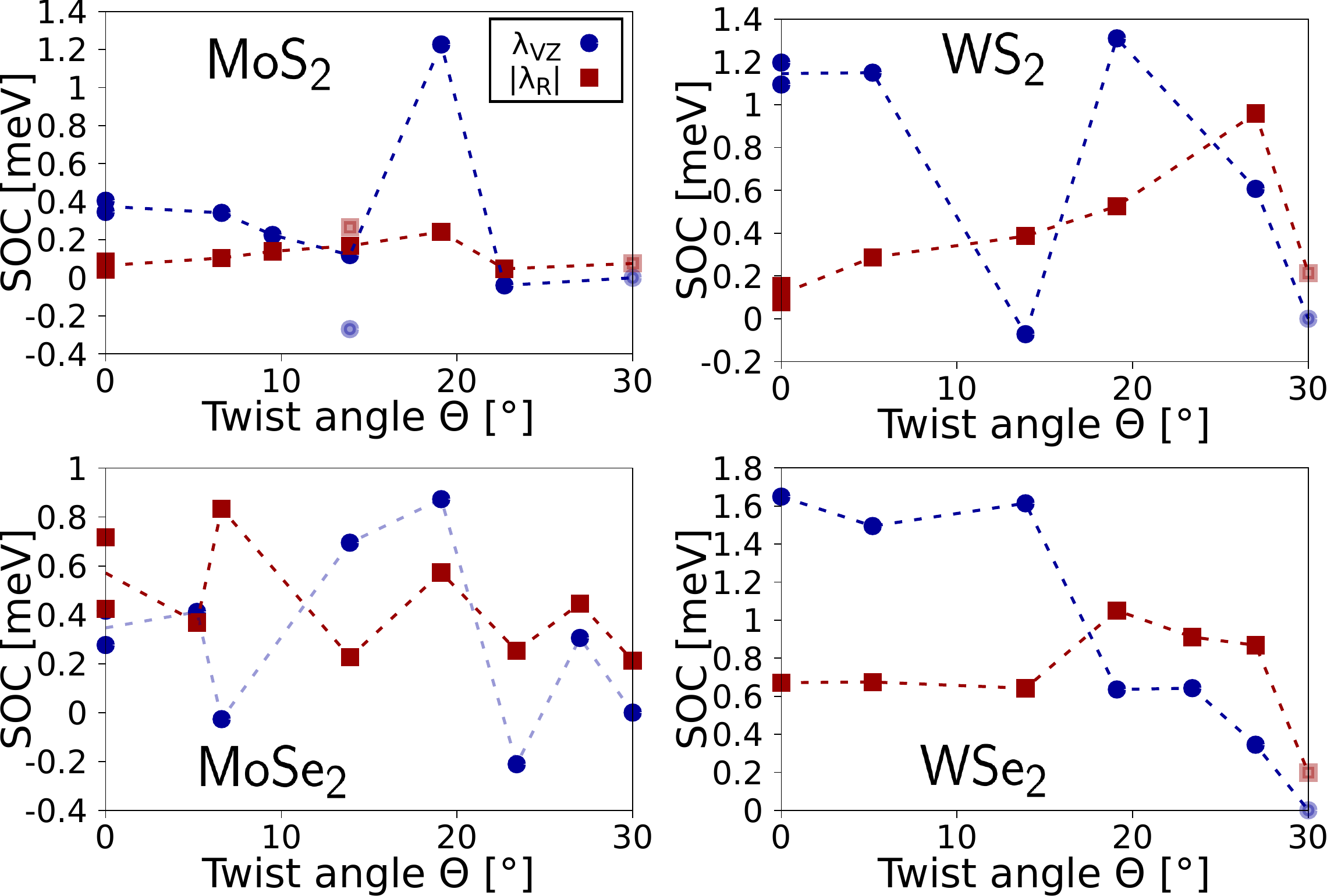}
     \caption{The uncorrected spin-orbit parameters $\lambda_{\text{R}}$ (red squares) and $\lambda_{\text{VZ}}$ (blue dots) of the graphene Dirac cone as function of the twist angle $\Theta$ for the four TMDCs MoS$_2$,  MoSe$_2$,  WSe$_2$ and  WS$_2$ similar to Fig.~\ref{Fig:params}. The transparent data points come from supercells with built-in strain $\epsilon>10\%$ and are therefore less reliable. We removed all data points with negative band offsets. The dotted line is a guide for the eyes connecting reliable data points. \label{Fig:paramsnoE}}
    \end{figure}

To correct the band offsets of our DFT calculations, we use electric fields in the $z$-direction. This is important to obtain the right spin-orbit parameters, since the energetic vicinity of the Dirac cone to the valence or the conductance band of the TMDC can massively influence the effectivity of the proximity effects. In Fig.~\ref{Fig:Efield} we depict this effect by exposing the 19.1\degree~MoS$_2$ supercell to different electric fields. Without any electric field, the Dirac cone of this supercell lies very close to the TMDC conductance band and therefore has very strong proximity SOC. By applying a electric field, we can bring the Dirac cone down in energy and exponentially reduce both the Rashba spin orbit-coupling $\lambda_{\text{R}}$ and the valley-Zeeman SOC $\lambda_{\text{VZ}}$.
In Fig.~\ref{Fig:paramsnoE} we show how Fig.~\ref{Fig:params} would look without the electric field corrections.



\bibliography{references}

\end{document}